\documentclass{elsart1p}
\usepackage{graphicx}
\usepackage{amssymb}
\begin{document}
\begin{frontmatter}
%
%
%
%
%
\title{Results from the STAR Beam Energy Scan Program}
%
%

\author{Lokesh Kumar (for the STAR Collaboration)}

\address{Department of Physics, Kent State University, USA}

\begin{abstract}
The main aim of 
the beam energy scan (BES) program at the Relativistic Heavy-Ion Collider (RHIC)
is to explore the quantum chromodynamics (QCD) 
phase diagram. 
The specific physics goal is to search for the phase boundary and the QCD
critical point. 
We present results from Au+Au collisions at various energies 
collected in the BES 
program by the Solenoidal Tracker At RHIC (STAR) experiment. 
First results on transverse momentum ($p_{T}$) spectra, $dN/dy$,
and average transverse mass ($\langle m_{T} \rangle$) for
identified hadrons produced at mid-rapidity for
$\sqrt{s_{NN}}$ = 7.7~GeV are presented. Centrality dependence
of $dN/dy$ and $\langle p_{T} \rangle$ are also discussed and
compared to corresponding data from other energies.
In addition, first results on charged hadron directed ($v_{1}$)
and elliptic flow ($v_{2}$) for 
$\sqrt{s_{NN}}=$ 7.7, 11.5, and 39 GeV are presented. 
New results on event-by-event fluctuations (particle ratio, net-proton
and net-charge higher moments) are presented for $\sqrt{s_{NN}}=$ 39 GeV.

\end{abstract}

%

%

\end{frontmatter}

\section{Introduction}
\label{}
The Relativistic Heavy-Ion Collider 
at Brookhaven National Laboratory (BNL) 
is built to study
the properties of a new state of matter, called Quark Gluon Plasma
(QGP). 
One of the major goals of heavy-ion collision experiments is to 
explore the QCD phase diagram.
The QCD phase diagram consists mainly two phases - 
the QGP phase, where the relevant degrees of freedom 
are quarks and gluons, and the hadronic phase.
Finite temperature
lattice QCD calculations~\cite{lattice}
at baryon chemical potential
$\mu_B=$ 0 suggest a cross-over above a critical temperature $T_c$
$\sim 170-190$ MeV 
from the hadronic to the QGP phase.
At large $\mu_B$, several 
QCD based calculations~\cite{firstorder} 
predict the quark-hadron phase transition to be 
of the first order. The point in the QCD phase plane ($T$ vs. $\mu_B$)
where the first order phase transition ends is the QCD critical point.
The BES program at RHIC~\cite{9gevprc,bmqm2010} aims to 
search the QCD phase boundary and 
QCD critical point.  
The QCD phase diagram can be accessed by varying temperature $T$ and 
baryonic chemical potential $\mu_{\rm{B}}$. Experimentally this can be achieved 
by varying the colliding beam energy.
The STAR took data in the year 2010 
for the beam energies $\sqrt{s_{NN}}=$ 7.7 GeV, 11.5 GeV, and 39 GeV 
as a first phase of the BES program.
\begin{table}
\begin{center}
\caption{\label{table1}
The BES energies, corresponding $\mu_{\rm{B}}$ values~\cite{cleymans}, and
total events with a minimum bias (MB) trigger collected by STAR during the BES run in 2010.}
\vspace{0.18cm}
\begin{tabular}{|c|c|c|}
\hline
$\sqrt{s_{NN}}$ (GeV) & $\mu_{\rm{B}}$ (MeV) & Events (Million MB)\\[0.01mm]
\hline
7.7           & 410 & 5 \\ [0.001mm]
\hline
11.5          & 300 & 7.5 \\ [0.001mm]
\hline
39            & 112 & 250 \\ [0.001mm]
\hline
\end{tabular}
\end{center}
\end{table}
\begin{figure}
\begin{center}
\includegraphics[scale=0.31]{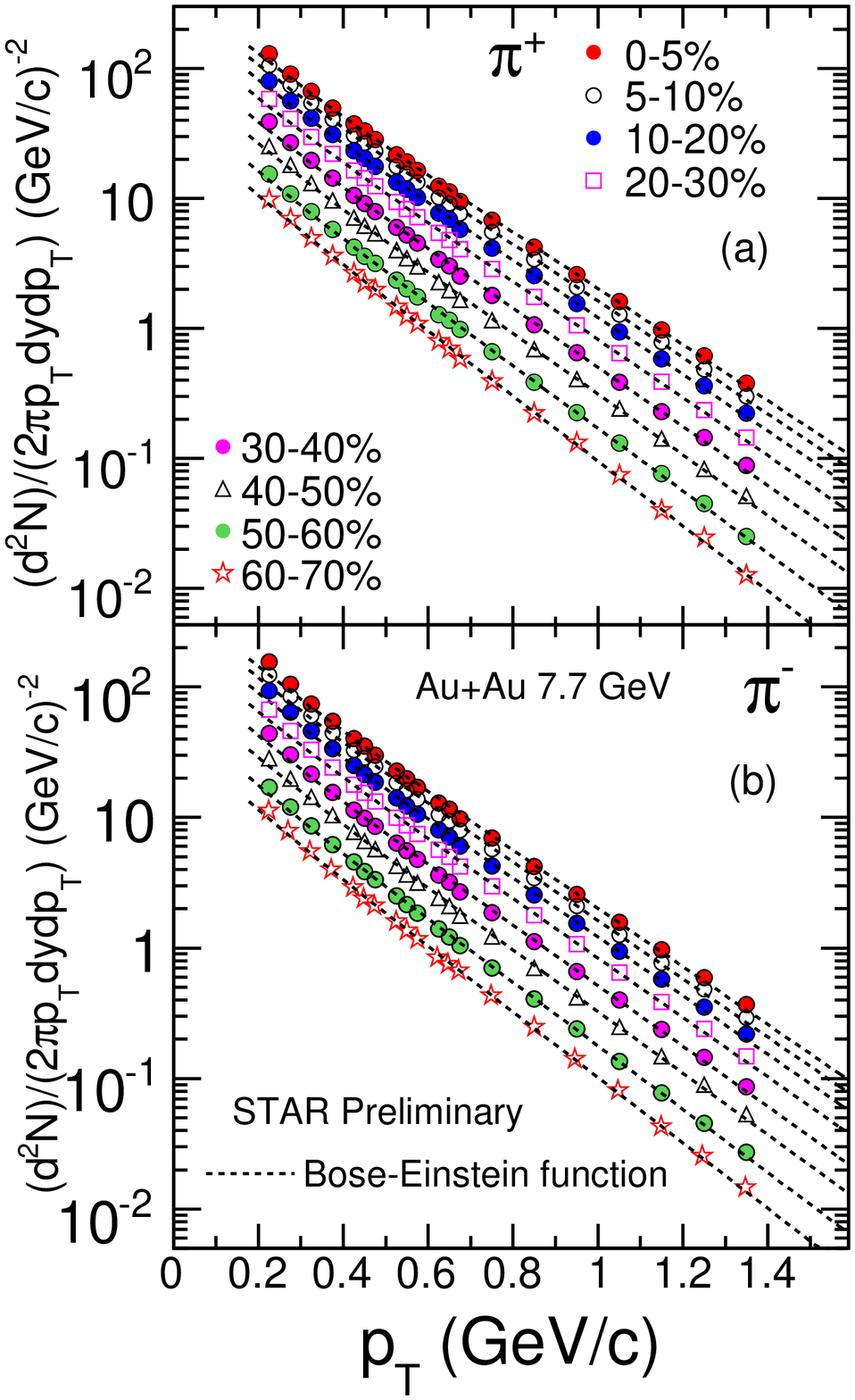}
\includegraphics[scale=0.33]{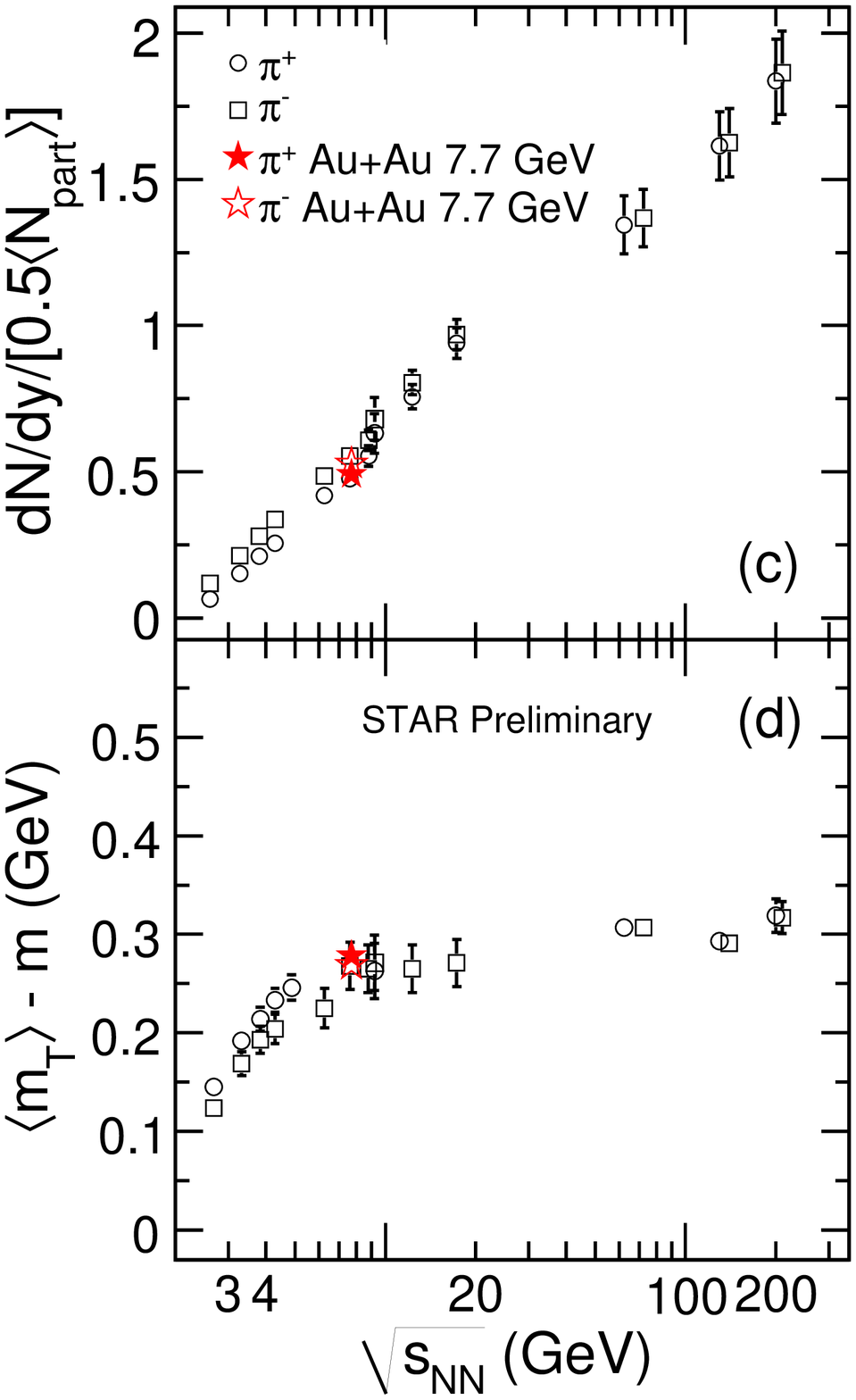}
\caption{(color online) Left panel: 
Transverse momentum spectra for charged pions at mid-rapidity 
($|y|<0.1$) in Au+Au collisions at $\sqrt{s_{NN}}=$ 7.7 GeV. The lines are the Bose-Einstein
fits to the distributions. 
Right panel: 
(c) $dN/dy$ normalized by 
$\langle N_{\mathrm {part}} \rangle$/2 and 
(d) $\langle m_{T} \rangle - m$ of $\pi^{\pm}$, plotted as a function
of collision energy. 
See text for details.
The errors shown are the quadratic sum of statistical and systematic uncertainties,
except for $\sqrt{s_{NN}}=$ 7.7 GeV, which has only statistical errors.
}
\label{spectra_dndy}
\end{center}
\end{figure}

The results presented here are based on data taken at STAR~\cite{star} 
for Au+Au collisions at $\sqrt{s_{NN}}$ = 7.7, 11.5, and 39 GeV in the year 2010. 
The main detector subsystem used for particle identification is the 
Time Projection Chamber (TPC)~\cite{tpc}.
Particle identification is enhanced up to higher $p_{T}$ with the recent inclusion 
of full barrel
Time Of Flight (TOF)~\cite{tof} detector.
The raw yields are extracted
at low-$p_{T}$ using ionization energy loss ($dE/dx$) from TPC, and at higher $p_{T}$ using 
TOF information. The identified particle results are presented for the mid-rapidity $|y|<0.1$
region. Directed flow results are obtained using the Beam Beam Counter 
(BBC) which provides event plane determination
at forward pseudorapidities
 (3.8 $\le |\eta| \le 5.2$).
Event planes used
for the elliptic flow results are provided by both TPC ($|\eta| < 1$) and 
Forward Time Projection Chamber (FTPC)~\cite{ftpc} (2.5 $\le |\eta| \le 4.2$). 
Table~\ref{table1} lists the BES energies, corresponding $\mu_{\rm{B}}$ values, and
total events collected by STAR during the BES program in the year 2010.

\section{Results}
\begin{figure}
\begin{center}
\includegraphics[scale=0.28]{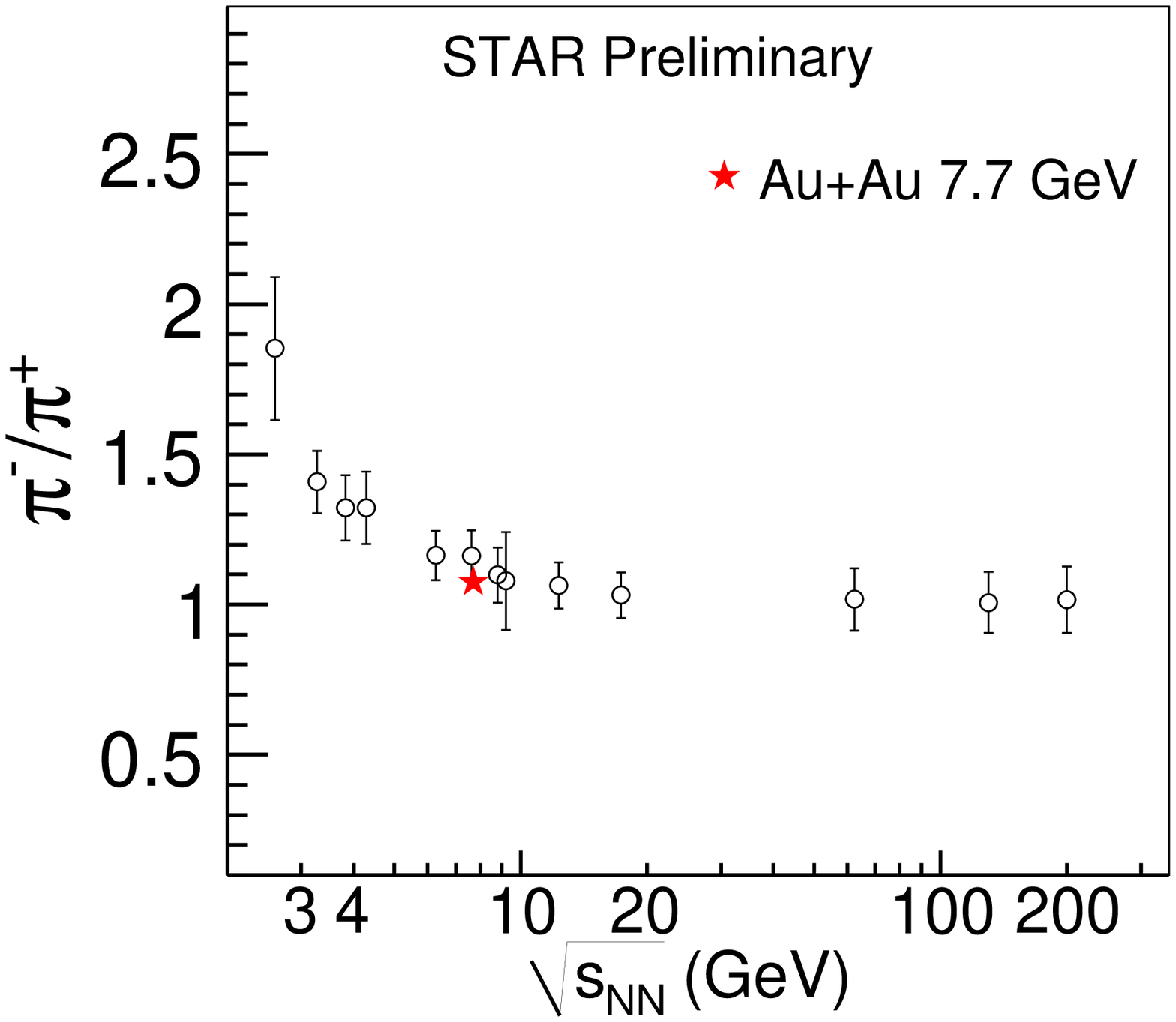}
\includegraphics[scale=0.34]{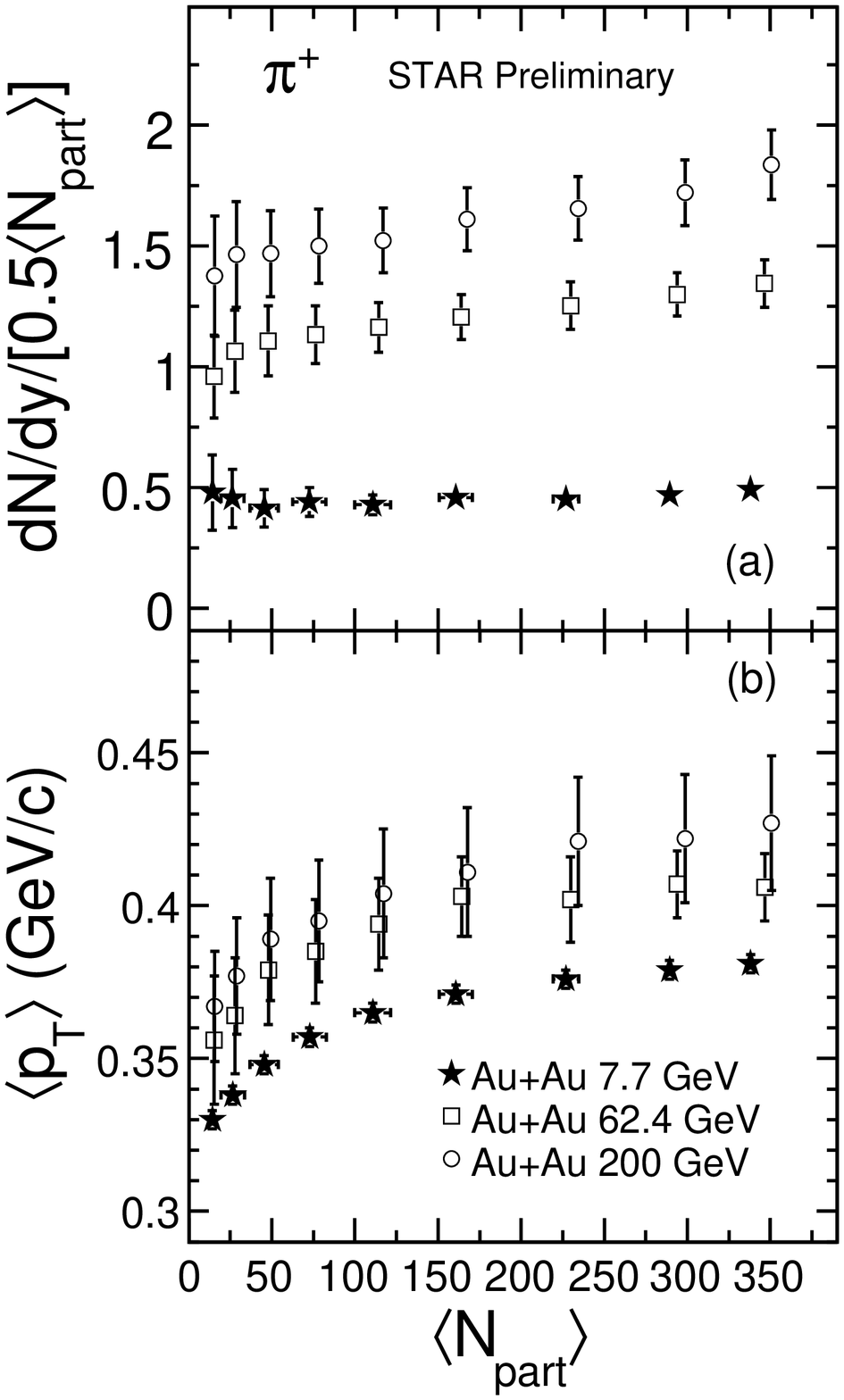}
\caption{
Left panel (color online): 
$\pi^{-}/\pi^{+}$  ratio at mid-rapidity 
($|y|< 0.1$) for central 0--5\% Au+Au collisions at 
$\sqrt{s_{NN}}$ = 7.7 GeV compared to previous results 
from AGS~\cite{ags}, SPS~\cite{sps}, and 
RHIC~\cite{9gevprc,STARPID}.
Right Panel: (a) $dN/dy$ normalized by $\langle N_{\mathrm {part}} \rangle / 2$ and 
(b) $\langle p_{T} \rangle$ of $\pi^{+}$,
plotted as a function 
of $\langle N_{\mathrm {part}} \rangle$. 
The errors shown are quadratic sum of statistical and systematic uncertainties,
except for $\sqrt{s_{NN}}=$ 7.7 GeV, which has only statistical errors.}
\label{rat_dndy}
\end{center}
\end{figure}
\subsection{Transverse momentum spectra at $\sqrt{s_{NN}}=$ 7.7 GeV}
Figure~\ref{spectra_dndy} (left panel) shows the transverse momentum spectra 
for $\pi^{\pm}$ in Au+Au collisions at $\sqrt{s_{NN}}$ = 7.7 GeV. 
The results are shown for various collision centrality classes as listed in the
figure. 
The pion
spectra presented here have been corrected for the weak decay feed-down and
muon contamination.
The particle production can be characterized 
by studying the $dN/dy$ and $\langle m_{T} \rangle - m$ for the 
produced hadrons, where $m$ is the mass of the hadron and 
$m_{T}$ = $\sqrt{m^{2} + p_{T}^{2}}$ is its transverse mass. These are discussed
in the next section.

\subsection{Energy dependence of yield, $\langle m_{T} \rangle$, and anti-particle to particle ratio}
Figure~\ref{spectra_dndy} (c) shows $dN/dy$ normalized 
by $\langle N_{\mathrm {part}} \rangle$/2 for $\pi^{\pm}$ 
in 0--5\% central Au+Au collisions 
at $\sqrt{s_{NN}}$ = 7.7 GeV and are compared
to previous results at AGS~\cite{ags}, SPS~\cite{sps}, and 
RHIC~\cite{9gevprc,STARPID}. Within errors,
the yields are consistent
with previous results at similar $\sqrt{s_{NN}}$.
The $\pi^{-}/\pi^{+}$ ratio at $\sqrt{s_{NN}}$ = 7.7 GeV is around 1.1
(Fig.~\ref{rat_dndy} left panel).
Figure~\ref{spectra_dndy}~(d) shows the 
$\langle m_{T} \rangle - m$ for $\pi^{\pm}$ in 
0--5\% central Au+Au collisions at $\sqrt{s_{NN}} = $ 7.7 GeV. The results are 
also compared to previous measurements at AGS~\cite{ags}, SPS~\cite{sps}, and 
RHIC~\cite{9gevprc,STARPID}.
The results from  Au+Au collisions at $\sqrt{s_{NN}}$ = 7.7 GeV are 
consistent with corresponding measurements at SPS energies at 
similar $\sqrt{s_{NN}}$. Both $dN/dy$ and $\langle m_{T} \rangle - m$
are obtained using data in the measured $p_{T}$ ranges and 
extrapolating 
using a Bose-Einstein functional form for the unmeasured $p_{T}$ ranges.
For the present mid-rapidity measurements, the 
contribution to the yields from extrapolation to the total yield is about 30\% for $\pi^{\pm}$.

The $\langle m_{T}\rangle - m$
values increase with $\sqrt{s_{NN}}$ at lower AGS energies, stay
independent of $\sqrt{s_{NN}}$ at the SPS and RHIC 7.7 GeV collisions, 
then tend to somewhat rise further with 
increasing $\sqrt{s_{NN}}$ at the higher beam energies at RHIC. For a thermodynamic 
system, $\langle m_{T}\rangle - m$ can be an approximate representation of
the temperature of the system, and  $dN/dy$ $\propto$ $\ln(\sqrt{s_{NN}})$ 
may represent the
entropy. In such a scenario, these observations could reflect the characteristic 
signature of a first order 
phase transition, as proposed by Van Hove~\cite{vanhove}. 
However, there could be several other effects to which $\langle m_{T}\rangle - m$ is 
sensitive,
which also need to be understood for proper 
interpretation of the data~\cite{bedanga}. 

\subsection{Centrality dependence of $dN/dy$ and $\langle p_{T}\rangle$}
Figure~\ref{rat_dndy} (right panel) shows (a) the comparison of collision centrality 
dependence of 
$dN/dy$ of $\pi^{+}$ normalized by $\langle N_{\mathrm {part}} \rangle / 2$,
between new results at $\sqrt{s_{NN}}$ = 7.7 GeV and previously published results 
at $\sqrt{s_{NN}}$ = 62.4 and 200 GeV from the STAR 
experiment~\cite{STARPID}. 
The yields of charged pions decrease with decreasing collision energy.
At low energy, 
$dN/dy/[0.5\langle N_{\mathrm {part}} \rangle]$ for charged pions is almost
constant as a function of collision centrality.
This supports the idea that particle production is dominated by soft processes 
at 7.7 GeV.
Right panel (b) of Fig.~\ref{rat_dndy} shows the comparison of $\langle p_{T} \rangle$
as a function of $\langle N_{\mathrm {part}} \rangle$ for $\pi^{+}$
from 
Au+Au collisions  at $\sqrt{s_{NN}}$ = 7.7 GeV with the same from collisions 
at  $\sqrt{s_{NN}}$ = 62.4 and 
200 GeV~\cite{STARPID}.
For the collision centralities studied, the dependencies of $\langle p_{T} \rangle$
on $\langle N_{\mathrm {part}} \rangle$ at $\sqrt{s_{NN}}$ = 7.7 GeV
are similar to those at $\sqrt{s_{NN}} = $ 62.4 and 200 GeV. 
The values of $\langle p_{T} \rangle$ increase from peripheral to central collisions.
This indicates that collectivity increases with collision centrality.
The $\langle p_{T} \rangle$ values also increase with collision energy.
\begin{figure}
\begin{center}
\includegraphics[scale=0.31]{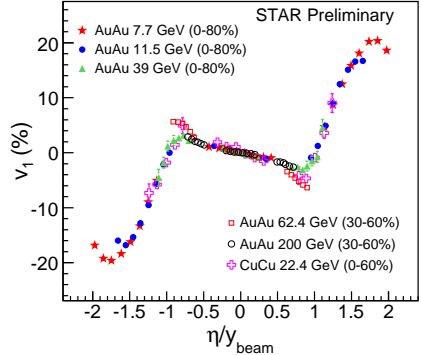}
\caption{
(color online)
 Charged hadron $v_{1}$ vs. $\eta/y_{\rm{beam}}$.
The errors shown are statistical.
 }
\label{v1}
\end{center}
\end{figure}

\subsection{Azimuthal anisotropy}
There are two types of 
azimuthal anisotropy that are widely studied in heavy-ion collisions, 
directed flow $v_{1}$ and elliptic flow $v_{2}$. 
Directed flow measurements at forward rapidities describe
the ``side-splash'' motion of the collision products.
The dependence of 
$v_1$ on $\eta$ around mid-rapidity 
is discussed in the literature 
as a possible signature of a first order
phase transition~\cite{directed}. 
Elliptic flow provides the possibility to gain information about the degree of thermalization
of the hot, dense medium. 
\begin{figure}
\begin{center}
\includegraphics[scale=0.31]{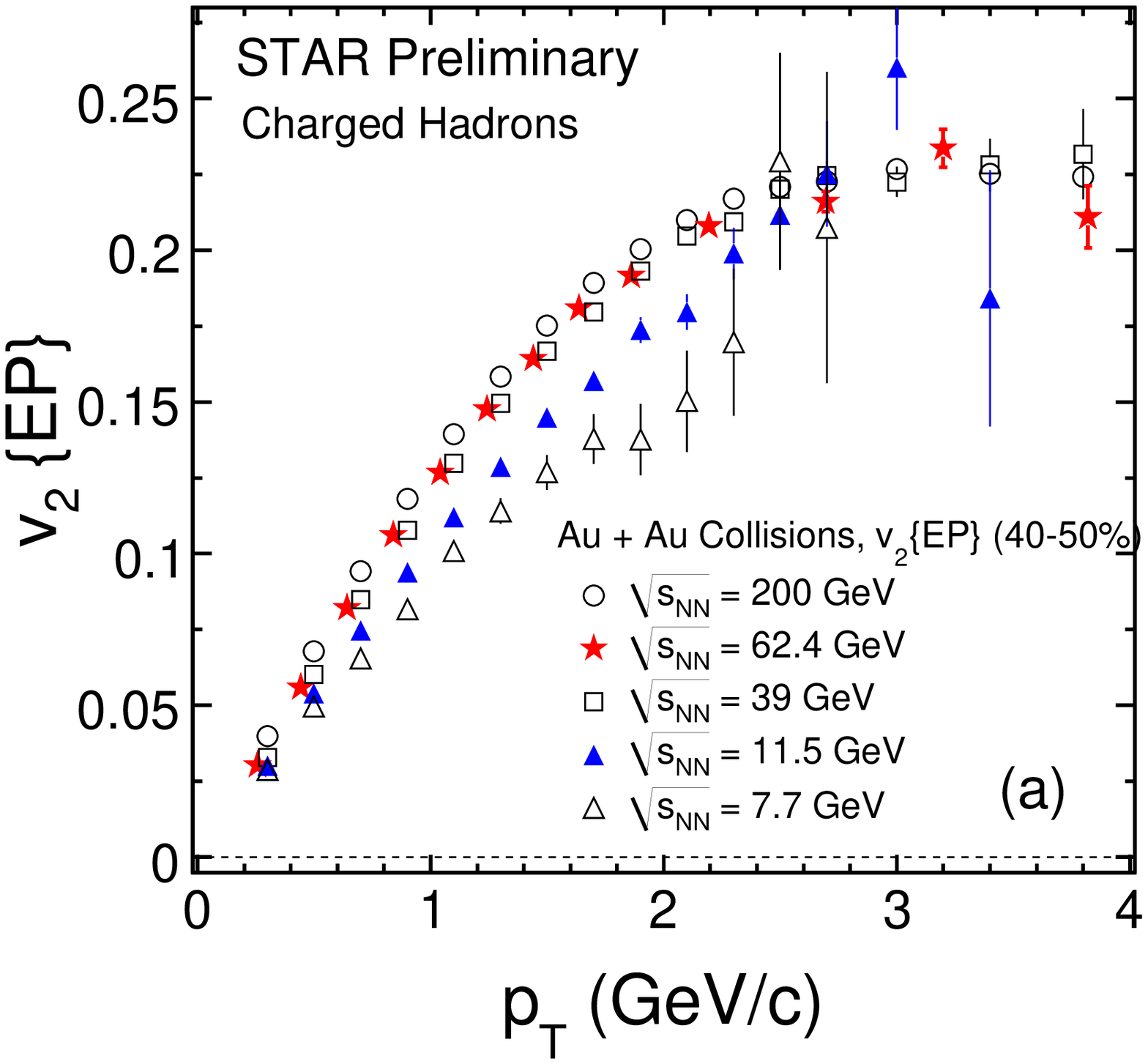}
\includegraphics[scale=0.32]{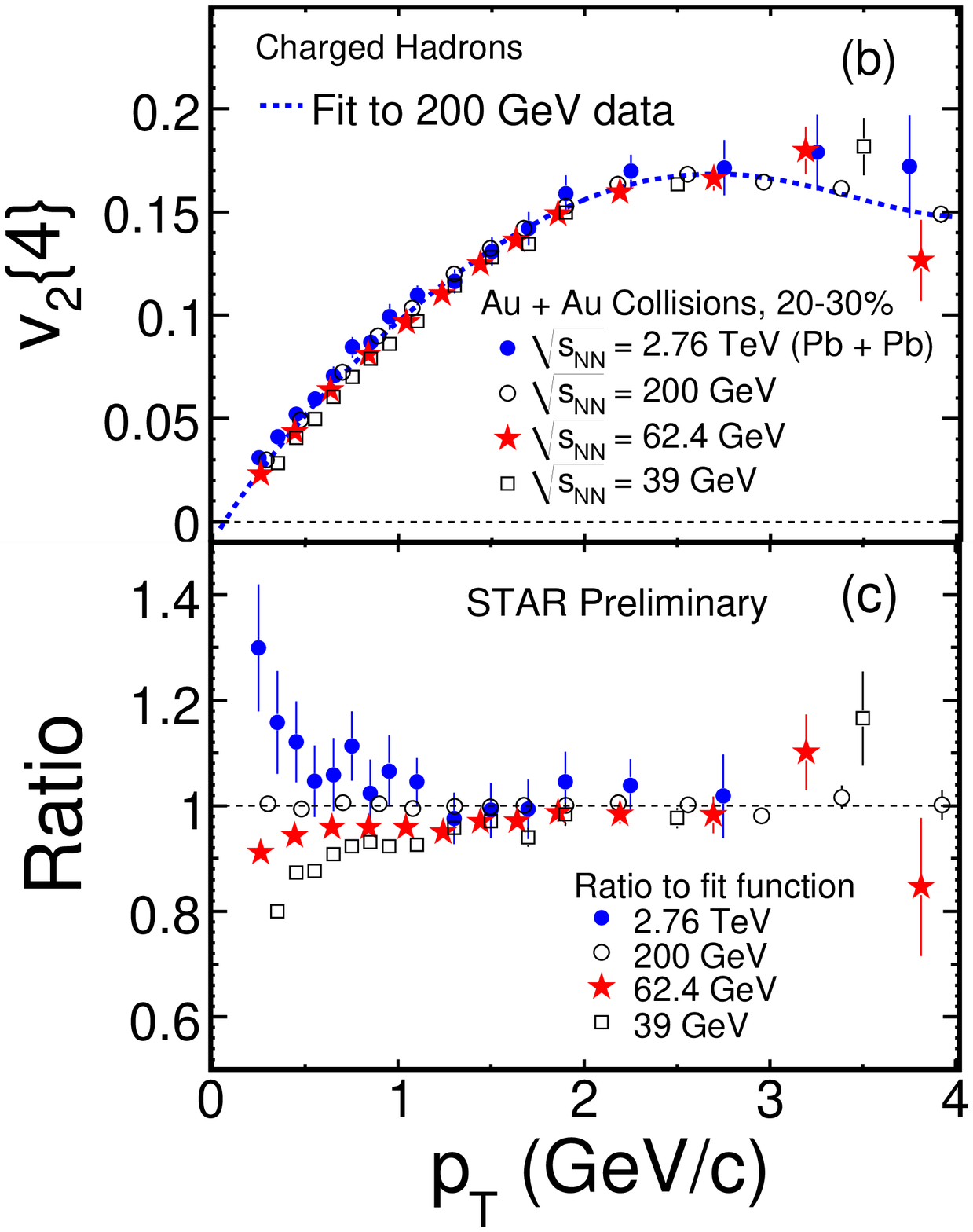}
\caption{(color online) Left panel:
(a) $v_{2}$ obtained from event plane method plotted as a function of $p_T$ for charged
hadrons.
The error bars include only statistical uncertainties.
Right panel: (b) $v_{2}$ obtained using the 4-particle cumulant method 
plotted as a
function of $p_T$ for charged hadrons.
The line represents the fit to 200 GeV data. (c) Ratio of the fit in 
(b) to $v_{2}$ from all the energies as a function of $p_T$.
}
\label{v2}
\end{center}
\end{figure}

Figure~\ref{v1} shows charged hadron $v_{1}$ 
results for the 0--80\% 
Au+Au collisions
at $\sqrt{s_{NN}}$~=~7.7 GeV, 11.5 GeV, and 39 GeV.
These are compared to
corresponding results from 30--60\% 
Au+Au collisions at $\sqrt{s_{NN}}$ = 62.4,
200 GeV~\cite{v14systempaper}, and 
0--60\% 
Cu+Cu collisions 
at $\sqrt{s_{NN}} = $ 22.4 GeV~\cite{yp}.
The mid-rapidity region corresponds to the produced particles, 
while the forward rapidity corresponds to the transported particles. 
At mid-rapidity, all the results have comparable values. At
forward rapidity, the trend of $v_{1}$ 
is energy dependent~\cite{9gevprc}. 
When $|\eta|$ is scaled with the corresponding beam
rapidities ($y_{\rm{beam}}$) for different energies, a universal curve is
obtained as shown in the figure for 
the measured $|\eta|/y_{\rm{beam}} < 0.5$ range.
The $y_{\rm{beam}}$
for $\sqrt{s_{NN}}$ = 7.7, 11.5, 39, 22.4, 62.4, and 200 GeV are 
2.1, 2.5, 3.2, 3.7, 4.2, and 5.4 respectively.

\begin{figure}
\begin{center}
\includegraphics[scale=0.29]{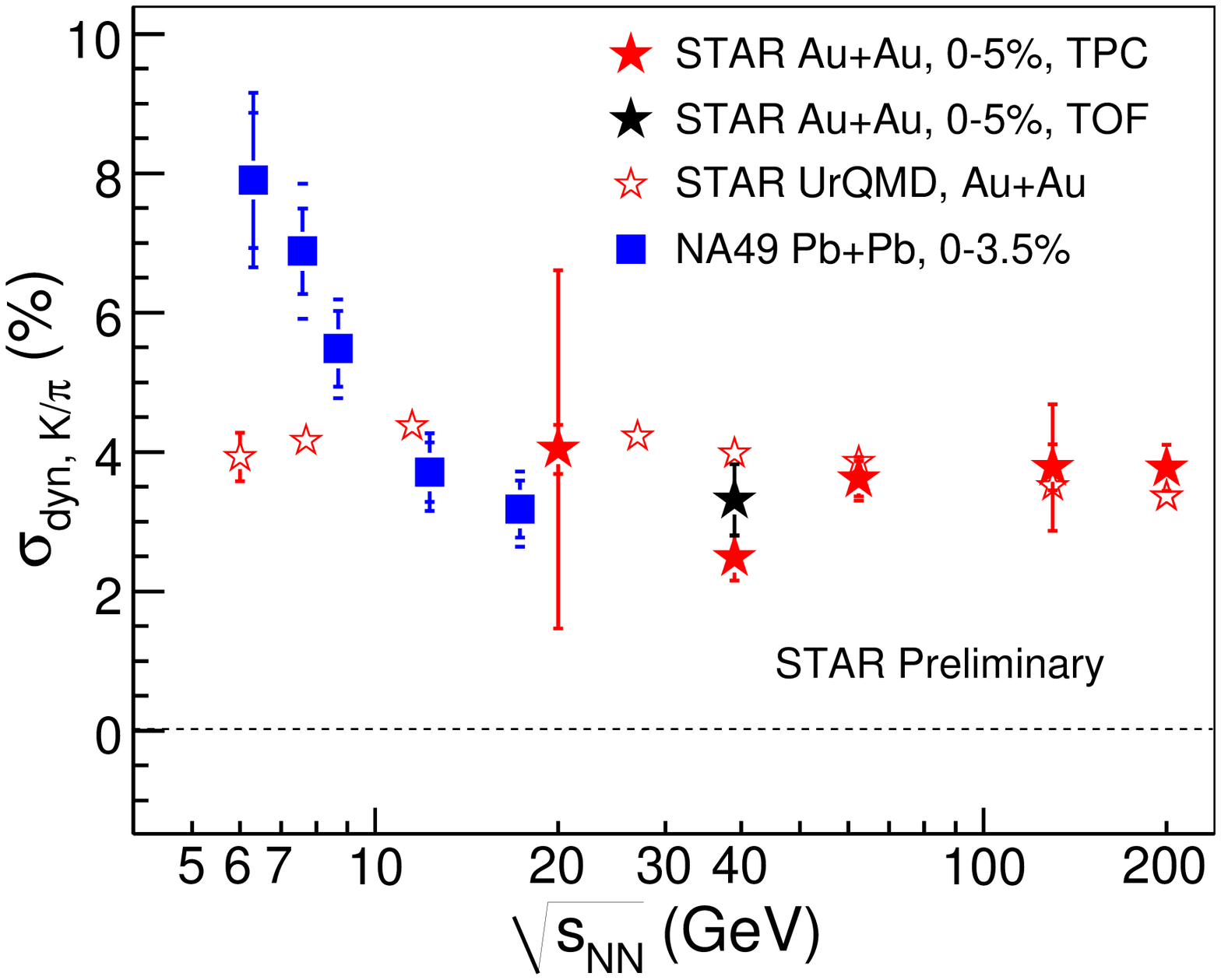}
\includegraphics[scale=0.29]{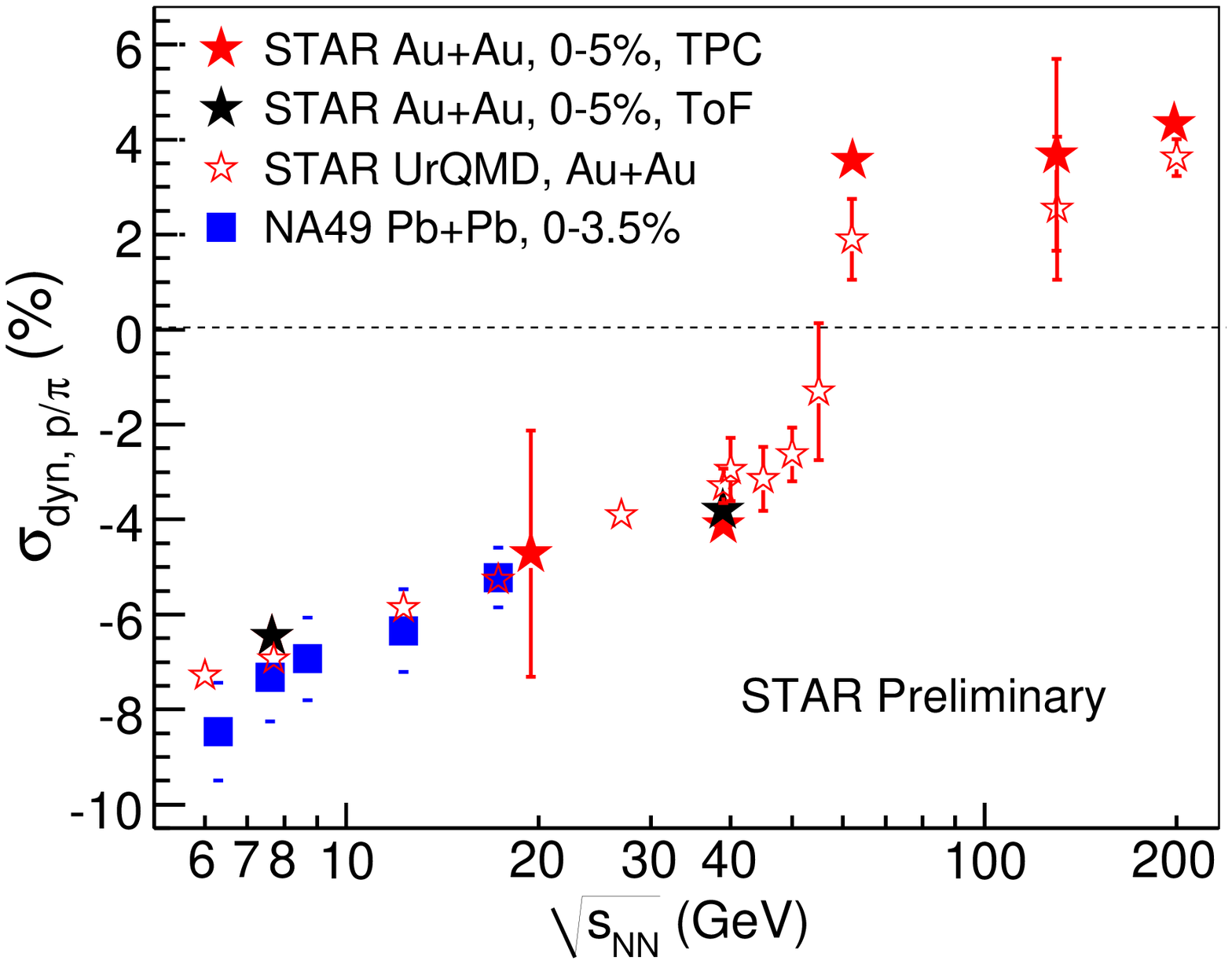}
\caption{ (color online) Left panel:
$K/\pi$ fluctuations expressed as $\sigma_{\rm{dyn}}$, plotted as a function of 
collision energy.
The errors
are quadratic sum of systematic and statistical uncertainties except 
for $\sqrt{s_{NN}} =$ 39 GeV,
which has only statistical errors. 
Right panel: same as (a) but for $p/\pi$ fluctuations.
}
\label{sig_dyn}
\end{center}
\end{figure}

Figure~\ref{v2} (a) shows $v_{2}$ obtained from event plane 
method~\cite{ep_meth}
plotted as a function of $p_T$ for charged
hadrons in 40--50\% Au+Au collisions at 
$\sqrt{s_{NN}} =$ 7.7, 11.5, and 39 GeV. The results are compared 
to corresponding results from $\sqrt{s_{NN}} =$ 62.4 and 200 GeV~\cite{starv2}.
The figure shows that $v_{2} (p_T)$
at 7.7 GeV is smaller than that at 11.5 GeV which is smaller than that at 39 GeV,
suggesting that elliptic flow increases 
as the BES energies increase.
Right panel (b) shows $v_{2}$ obtained using the 4-particle cumulant 
method~\cite{4p_cum} 
plotted as a
function of $p_T$ for charged hadrons in 20--30\% Au+Au collisions at 
$\sqrt{s_{NN}} =$ 39, 62.4, and 200 GeV, from STAR. The results are compared
with corresponding results in Pb+Pb collisions at 
$\sqrt{s_{NN}} =$ 2.76 TeV, from the ALICE experiment~\cite{alicev2}. The line
represents the fifth-order polynomial fit to 200 GeV data. Panel (c) shows 
the ratio of the fit to $v_{2}$
from all the energies as a function of $p_T$. It is observed that $v_{2}\{4\}(p_T)$
for all the energies show similar values (within $\sim 10\%$ from 200 GeV data) 
beyond $p_T=$ 500
MeV$/c$. 
This saturation is very interesting considering the wide energy range 39 GeV 
to 2.76 TeV.

\subsection{Fluctuations}
The non-monotonic behavior of fluctuations in 
the particle ratios such as $K/\pi$ and $p/\pi$ as a function of 
collision energy could indicate the presence of a QCD critical point or
the phase transition. 
These ratio
fluctuations are quantified by the variable $\nu_{\rm{dyn}}$~\cite{k2pi_paper}. 
Earlier measurements~\cite{sigdyn} of
particle ratio fluctuations used the variable given by:~
$\sigma_{\rm{dyn}}=\rm{sign}(\sigma^{2}_{\rm{data}}-\sigma^{2}_{\rm{mixed}})\sqrt{|\sigma^{2}_{\rm{data}}-\sigma^{2}_{\rm{mixed}}|}$,
where $\sigma$ is the relative width of the $K/\pi$ and $p/\pi$ distribution 
in either real data or mixed events. 
It has been shown that $\sigma_{\rm{dyn}}^{2}\approx\nu_{\rm{dyn}}$.

Figure~\ref{sig_dyn} (left panel) shows the energy dependence of 
dynamical $K/\pi$ fluctuations
expressed as $\sigma_{\rm{dyn}}$~\cite{terry}. 
The 0-3.5\% central Pb+Pb collisions NA49 results 
(solid squares) show decrease in 
dynamical fluctuations as a function of collision energy. The 0-5\% central Au+Au
collisions STAR results (solid stars) for $\sqrt{s_{NN}} =$ 19.6, 62.4, 130 and 200 GeV,
are constant as a function of beam energy. The newly measured $K/\pi$ 
fluctuation results from 0-5\% central Au+Au collisions at 
$\sqrt{s_{NN}} =$ 39 GeV are similar to other
STAR results~\cite{k2pi_paper}. The UrQMD model calculations  (open stars) using 
the STAR detector
acceptance, give approximately 4\% $K/\pi$ fluctuations for all the beam energies.
The right panel of Fig.~\ref{sig_dyn} shows the energy dependence of dynamical $p/\pi$ 
fluctuations
expressed as $\sigma_{\rm{dyn}}$~\cite{terry}. The STAR results (solid stars) 
from 0-5\% central Au+Au 
collisions 
are compared with those from 0-3.5\% central Pb+Pb collisions from the 
NA49 experiment 
(solid squares). The new results from 0-5\% central Au+Au collisions 
at $\sqrt{s_{NN}} =$ 7.7 and 39 GeV are also shown. All the data points show increase 
in $p/\pi$ fluctuations as
a function of beam energy. The STAR results from 7.7 GeV are in close agreement with
those from the NA49 experiment at similar $\sqrt{s_{NN}}$. The UrQMD model calculations
(open stars) 
after correcting for the STAR detector acceptance reproduce the increasing trend seen in 
data as a function of beam energy. 
The STAR results presented here are from the similar $p_T$ ranges 
(pion, kaon: 0.2--0.6 GeV/$c$ and proton: 0.4--1.0 GeV/$c$) 
for both TPC and TOF.

It has been shown that higher moments of distributions of conserved quantities 
(net-baryon number, net-strangeness and net-charge), measuring deviations 
from a Gaussian, are sensitive to the critical point fluctuations~\cite{cp_mom}. 
The moments, standard deviation $\sigma$, skewness $S$, and kurtosis $\kappa$,
of conserved quantities distributions, are defined as:
$\sigma=\sqrt{\langle (N-\langle N \rangle)^{2} \rangle}$, 
$S=\langle (N-\langle N \rangle)^{3} \rangle/\sigma^{3}$, and
$\kappa=\langle (N-\langle N \rangle)^{4} \rangle/\sigma^{4}-3$, respectively.
The products of the moments such as $\kappa\sigma^{2}$ and $S\sigma$, are related to
the ratio of conserved quantities number susceptibilities ($\chi$) at a given 
temperature ($T$) computed
in QCD models as : $S\sigma\sim\chi^{(3)}/\chi^{(2)}$ and 
$\kappa\sigma^{2}\sim\chi^{(4)}/\chi^{(2)}$. 
Close to a critical point, models predict the 
conserved quantities number distributions
to be non-Gaussian and susceptibilities to diverge causing $S\sigma$ and $\kappa\sigma^{2}$
to deviate from constants and have large values.

\begin{figure}
\begin{center}
\includegraphics[scale=0.35]{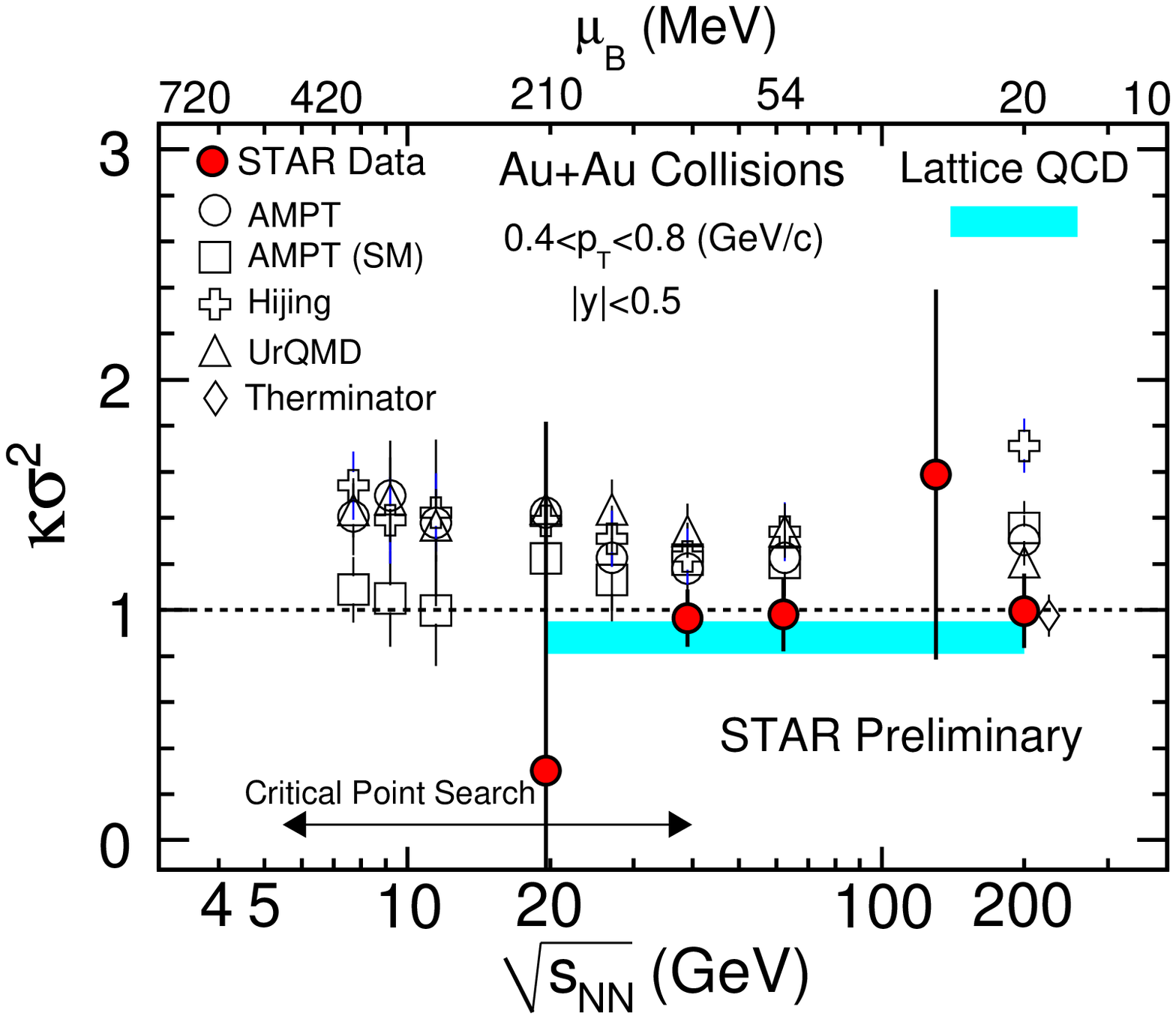}
\includegraphics[scale=0.31]{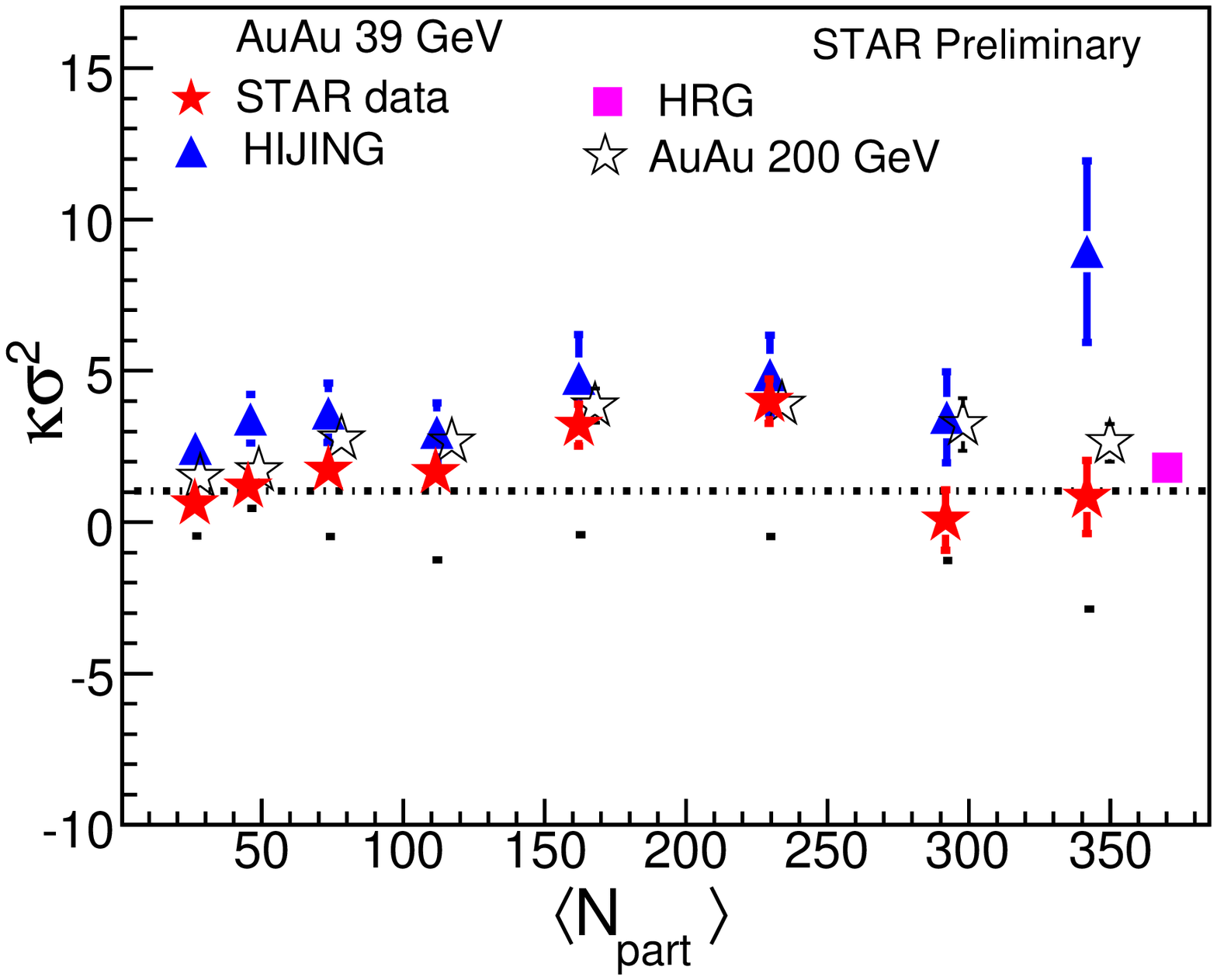}
\caption{ (color online) Left panel:
$\sqrt{s_{NN}}$ dependence of $\kappa\sigma^{2}$ for net-proton distributions 
measured
at RHIC. 
Errors are quadratic sum of
systematic and statistical uncertainties, except for $\sqrt{s_{NN}}=$ 39 GeV, which has
only statistical errors.
Right panel: $\kappa\sigma^{2}$ for net-charge distributions as a function of $\langle N_{\mathrm {part}} \rangle$.
Results from $\sqrt{s_{NN}}=$ 39 GeV are compared with those from $\sqrt{s_{NN}}=$ 200 GeV~\cite{nc_200}. Also shown
are the calculations from the HIJING and HRG models.
}
\label{kurt}
\end{center}
\end{figure}
Figure~\ref{kurt} (left panel) shows the energy dependence of $\kappa\sigma^{2}$ for 
net-proton distributions,
compared to several model calculations that do not include a critical point. 
The $\kappa\sigma^{2}$ as a 
function
of the beam energy does not show any non-monotonic behavior 
and is consistent with unity. The new STAR 
result for $\kappa\sigma^{2}$ from $\sqrt{s_{NN}}=$ 39 GeV is 
consistent with previous STAR
measurements~\cite{mom_pap}.
The right panel shows the $\kappa\sigma^{2}$ for net-charge 
distributions as a function of $\langle N_{\mathrm {part}} \rangle$.
STAR results for $\sqrt{s_{NN}}=$ 39 and 200 GeV are consistent with each other. HIJING results
are consistent with the data except for the most central collisions. 
The $\kappa\sigma^{2}$ for central 
collisions is consistent with the hadron resonance gas (HRG) model~\cite{hrg} 
which assumes 
thermal equilibrium.

\section{Summary}
The RHIC beam energy scan program has started. The aim of the 
BES program is to search for
the QCD phase boundary and QCD critical point. In the first phase of the
BES program, the STAR experiment took data for the beam energies $\sqrt{s_{NN}}$ = 7.7, 11.5, and 39 GeV 
during 2010. 
The event statistics goals for three beam energies were achieved. 
The first 
results from the BES program are presented in this paper. Measurements of 
identified particle production 
at $\sqrt{s_{NN}}$ = 7.7 GeV suggest that particle production scales
with $\langle N_{\mathrm {part}} \rangle$.
The measurements of $\langle p_{T} \rangle$ as a function of 
collision centrality indicate that 
the collectivity increases with collision centrality. The mid-rapidity 
$\pi^{-}/\pi^{+}$ ratio 
at $\sqrt{s_{NN}}$ = 7.7 GeV is close to 1.1. Directed flow results for 
all the beam energies 
scale with $\eta/y_{\rm{beam}}$ 
(for measured $|\eta|/y_{\rm{beam}} < 0.5$ range), 
extending the already established 
scaling behavior down to 7.7 GeV. 
The $v_{2}\{4\} (p_T)$ shows saturation 
above $p_T=500$ MeV$/c$ for all
the beam energies from
39 GeV through 2.76 TeV. Particle
ratio fluctuations for the energies presented are consistent with the established trends. The
net-proton and net-charge results are consistent with the HRG model.

We wish to acknowledge the support from DOE.

\end{document}